# Depression as a disorder of distributional coding

Matthew Botvinick[1,2], Zeb Kurth-Nelson[1,2], Timothy Muller[3], Will Dabney[1]

[1]DeepMind, London, UK
[2]University College London
[3]University of Oxford

**Abstract**

Major depressive disorder persistently stands as a major public health problem. While some progress has been made toward effective treatments, the neural mechanisms that give rise to the disorder remain poorly understood. In this Perspective, we put forward a new theory of the pathophysiology of depression. More precisely, we spotlight three previously separate bodies of research, showing how they can be fit together into a previously overlooked larger picture. The first piece of the puzzle is provided by pathophysiology research implicating dopamine in depression. The second piece, coming from computational psychiatry, links depression with a special form of reinforcement learning. The third and final piece involves recent work at the intersection of artificial intelligence and basic neuroscience research, indicating that the brain may represent value using a distributional code. Fitting these three pieces together yields a new model of depression's pathophysiology, which spans circuit, systems, computational and behavioral levels, opening up new directions for research.

## Introduction

Major depressive disorder (MDD) is a leading cause of disability worldwide, affecting more than 300 million people each year [1, 2]. Despite a steadily growing knowledge base concerning the biological and psychological correlates of the disorder, there is as of yet no consensus concerning its core underlying mechanisms. The pathophysiology of MDD thus remains a pressing open problem.

In the present perspective, we propose a new theory of depression. Rather than introducing new experimental data, our approach is instead to reach into the existing literature, pointing out potential connections among a set of previously separate findings, and suggesting how they can be fit together, like pieces of a puzzle, to reveal a new picture. As a model of this approach, we have in mind canonical work from the 1990s, which established a connection between dopamine signalling and the reward-prediction error from reinforcement learning theory [3] (see Figure 2). In that case, the key insight involved bringing together information from disparate literatures or disciplines. In the present work, we take a similarly transdisciplinary approach.

As we shall detail, the first piece in this puzzle is pathophysiology research implicating the ventral tegmental area (VTA) in depression. The second piece is work from computational psychiatry, linking depression with a particular form of reinforcement learning. The third and unifying piece in the account we will propose derives from basic neuroscience research, in particular recent findings concerning the representation of value in the mesolimbic dopamine system, and focusing on the idea of *distributional coding* [4]. Fitting these pieces together reveals a larger picture, which links stress-induced disturbances in VTA function, through computational interpretations of dopamine function, to the clinical and behavioral phenomenology of depression.

In order to spell out the relevant details, the following discussion begins with the evidence concerning the VTA in depression. We then shift to computational psychiatry and the profile of reinforcement learning in MDD. Finally, we turn to distributional coding, showing how all three pieces can be fit together to obtain a new theory.

## Puzzle piece 1: VTA dysfunction in depression

Although numerous brain systems have been implicated in depression [5, 6], some of the strongest evidence centers on the mesolimbic dopamine system, and in particular its central nucleus, the VTA. Existing evidence concerning VTA dysfunction in MDD takes diverse forms, and has been the subject of a wide range of interpretations [7, 8]. However, a particularly compelling account has been proposed by Grace and colleagues [9, 10, 11, 12]. Their work, across a series of studies, has shown that under normal conditions a sizable proportion of VTA dopamine neurons are functionally silent (for related evidence see [13], but also [14]), due to tonic inhibition imposed via projections from the ventral pallidum (though also potentially reflecting inputs from lateral habenula [15]). Crucially, depression appears to involve a magnification of this suppressive effect. Under chronic mild stress (CMS) in rodents, a well validated model of MDD [16], a striking increase occurs in the proportion of VTA dopamine neurons that are functionally inactive, without gross changes in the firing rates or patterns in their active neighbors [9, 10, 11] (Figure 1). Further investigation has provided evidence that this suppressive effect of CMS on VTA population activity is mediated by upstream effects on the basolateral amygdala and on infralimbic prefrontal cortex, an analog of primate Brodmann area 25 [10, 11], and again also potentially reflecting changes in lateral habenula activity [15]. It has also been shown that counteracting the suppressive effect of CMS by optogenetic stimulation in VTA can prevent the emergence of depressive-like behaviors [17].

Although some contrasting findings have been obtained in other animal models (e.g., [18, 19]), the idea that depression may involve a dropout or silencing of a subpopulation of VTA neurons – the *dropout model*, as we shall call it – is consistent with experimental evidence showing that lesioning or optogenetic silencing of VTA dopamine neurons can induce behaviors resembling those induced by CMS [17, 20]. The dropout model also broadly coheres with two observations from human postmortem studies. The first is that VTA dopamine neurons can undergo degeneration in Parkinson's Disease, a disorder which is accompanied by depression in around 35% of cases [21, 22] (see also [23]). The second comes from research on late-life depression, where symptom severity has been found to vary inversely with the density of dopaminergic neurons within VTA [24]. In the dropout model, as proposed by Grace and colleagues [10], the loss of dopaminergic neurons in MDD is functional and reversible rather than permanent, but these postmortem studies nonetheless provide some convergent support for the idea that depression may stem from a silencing of dopaminergic neurons in VTA.

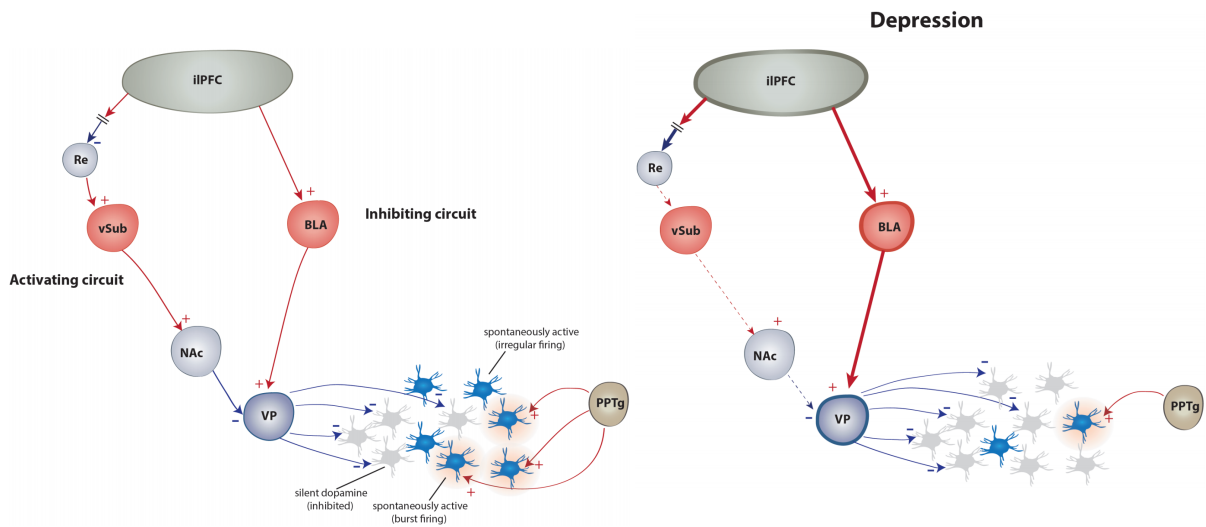

Figure 1: A series of studies by Grace and colleagues together indicate that chronic stress induces a suppression of activity in a subpopulation of VTA dopamine neurons, driven by augmented inhibitory input from the ventral pallidum (VP), itself the result of altered activity in upstream structuress including infralimbic cortex (ilPFC), nucleus reuniens of the thalamus (Re), ventral subiculum (vSub), nucleus accumbens (NAc), and basolateral amygdala (BLA). According to this model, reward-related input from the pedunculopontine tegmental nucleus (PPTg) triggers burst firing only in non-suppressed DA neurons. **Left**: Normal circuitry. **Right**: Alterations induced by chronic stress. From [10].

## Puzzle piece 2: Risk-sensitive RL

One appealing aspect of the dropout model is its broad fit with independent evidence indicating that dopamine plays a central role in motivation and valuation [25, 26, 27, 28], functions that are clearly disrupted in depression. However, if one consults contemporary research investigating what exact role dopamine plays in valuation and motivation, and then views the dropout model through the resulting lens, some quandaries arise.

As mentioned earlier, a large body of research has linked phasic dopamine signaling with the reward prediction error (RPE) from reinforcement learning theory (Figure 2). The RPE captures the difference between reward received from the environment, on the one hand, and an *a priori* prediction concerning that reward, on the other. The latter prediction, referred to as the *state-value estimate*, estimates a delay-weighted sum of all future rewards:

$$V(s_t) = \sum_{i=0:\infty} (\gamma^i r_{t+i}) \tag{1}$$

where $s_t$ and $r_t$ represent the state of the environment and the scalar reward received at time $t$, and $\gamma$ is a temporal discount parameter.[1]

In so-called *temporal difference learning*, the RPE ($\delta$) is computed on each time-step through a bootstrapping operation:

$$\delta = (r_t + \gamma V(s_{t+1})) - V(s_t) \tag{2}$$

The RPE is then used to adjust state-value estimates, bringing them gradually in line with experience, at a pace governed by a learning rate $\alpha$:

$$V_t \leftarrow \alpha \delta + V_t \tag{3}$$

In what has become the canonical account in neuroscience, phasic dopamine release from neurons in the VTA encodes the RPE, computed based on state-value representations encoded in the activities of other neurons both within and beyond the VTA [3]. Dopamine thus plays two intimately related roles: It signals reward-related surprise, and it drives the gradual updating, through learning, of value estimates.

At first blush, this account of dopaminergic function appears to fit well with the dropout theory of depression. If VTA dopamine neurons signal the RPE, and depression involves a reduction in the number of active VTA neurons, then one would expect neural measurements of RPE signalling to be attenuated in depressed patients. This is precisely what has been observed in numerous functional neuroimaging studies, where responses to reward have been measured in projection targets of the VTA, most notably the ventral striatum [32, 33, 34, 35] (see also [36, 37]). If RPE magnitude, as signaled by VTA population activity, drives value learning, then an additional prediction from the dropout model is that reward-driven learning should be impaired in MDD. Here again, empirical data align with predictions, as a large collection of studies have reported reduced reward-driven learning in depressed patients [32, 34, 38, 39, 40, 41, 42, 43, 44].

However, a closer look at the literature on reward-driven learning in depression reveals an interesting problem. In particular, a sizeable body of research indicates that, while learning from positive outcomes may be reduced in depression, learning from *negative* outcomes is less affected, and in some cases may be augmented [32, 34, 40, 41, 43, 45]. This observation is not predicted by any straightforward combination of the classical RPE account of dopamine with the dropout theory of depression. According to the former, outcomes that trigger a negative RPE should be associated with a dip in VTA firing rates below baseline, and thus a transient reduction in dopamine release in target structures like the striatum [3]. Importing this idea into the dropout model, the reduced population of active VTA dopamine neurons in depression should lead to a reduction in the amplitude of this dip, and thus an effective weakening of the learning signal. Why then is depression associated with a relative sparing of learning driven by negative outcomes?

---

[1] Those familiar with the RL literature will recognize a distinction between state values and action values (Q-values [29]). For simplicity, we focus on the former, but presume that it is obvious how the same account could be framed in terms of the latter.

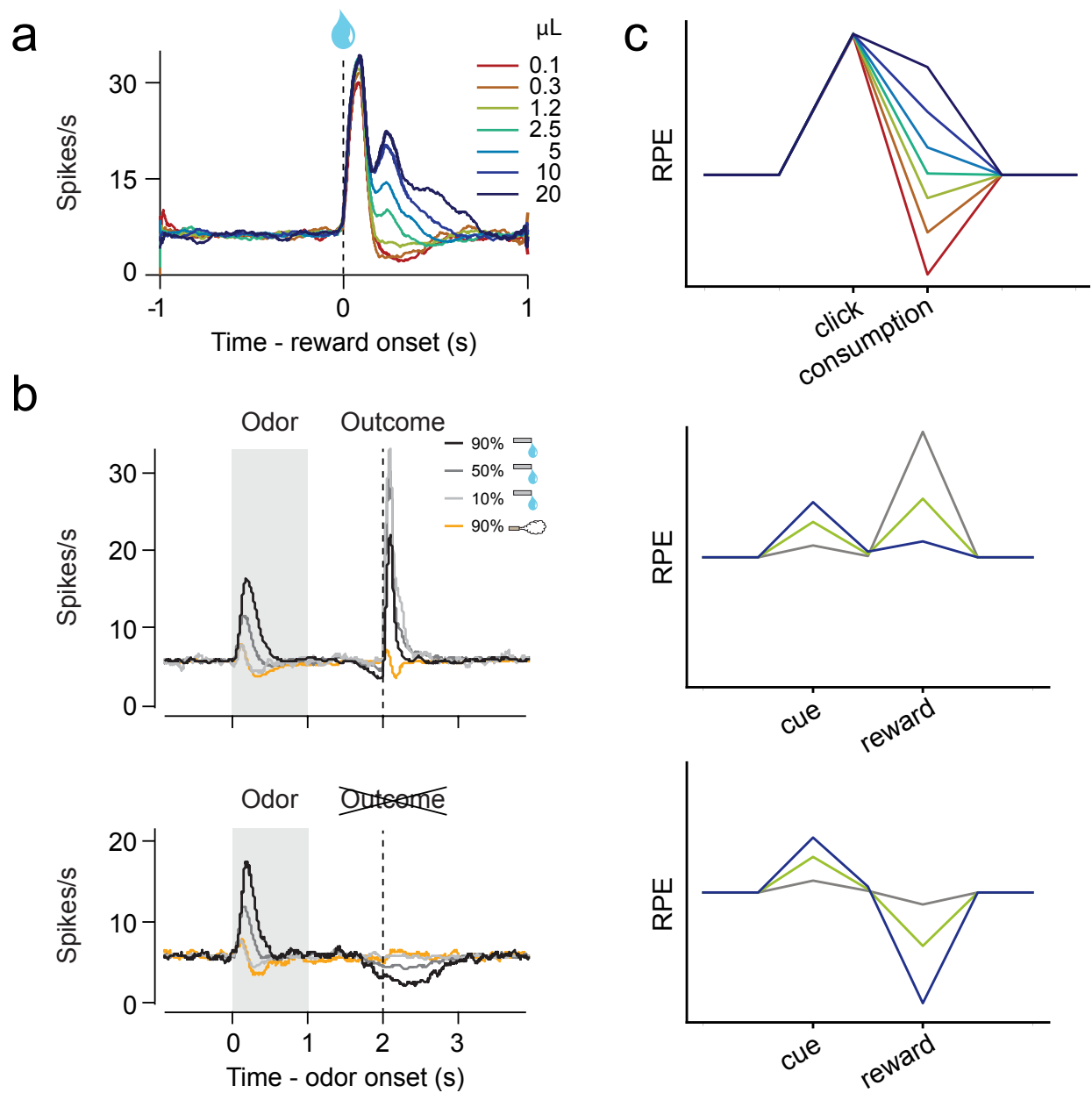

Figure 2: Phasic activity of dopaminergic neurons in the lateral VTA parallels the RPE from temporal difference RL. **a.** Firing rates, averaged across multiple neurons, in response to a juice reward, varying with the (unpredictable) delivered volume. Note the dip in activity for smaller-than-average volumes. From [30]. **b.** Average firing rates in response to a cue signalling the probability of a following appetitive (juice) or aversive (air-puff) event, with responses scaling with the probability of reward. Responses to subsequent stimulus delivery (top plot) or omission (below) scale with surprise. From [31]. **c.** RPE signals from simulations of the same tasks, using standard temporal difference RL (see Eq. 2-3).

A partial answer is offered by recent work from the field of computational psychiatry [46]. Here, Bennett and Niv [47] have suggested a connection between depression and a modified version of temporal difference learning called *risk-sensitive reinforcement learning*. Originally proposed in the machine learning literature [48] and already empirically tied to normative brain function [49], risk-sensitive RL elaborates on the standard temporal difference framework by introducing separate learning rates for positive and negative RPEs:

$$V(s_t) \leftarrow \begin{cases} \alpha^+ \delta + V(s_t) & \text{if } \delta > 0 \\ \alpha^- \delta + V(s_t) & \text{if } \delta < 0 \end{cases} \quad (4)$$

As noted by Bennett and Niv [47], this computational framework can capture the asymmetric disturbance of learning in depression if it is assumed that depression, by whatever mechanism, reduces the learning rate associated with positive outcomes ($\alpha^+$), while increasing or leaving unaltered the learning rate for negative outcomes ($\alpha^-$).

As Bennett and Niv [47] point out, this proposal also leads to another prediction about reward-based decision making in depression that turns out to be largely supported by empirical data. Specifically, the learning-rate asymmetry account predicts that depression should be associated with hightened risk aversion: In choice situations involving probabilistic outcomes, depressed individuals should behave as if they overestimate the probability of the less desirable possibilities. Consistent with this, depressed individuals consistently display increased risk aversion in self-report [50, 51, 52], laboratory tasks [53, 54, 55], and naturalistic settings [56, 57]. Risk aversion is tied to the phenomenon of 'avoidance' believed to play a key role in sustaining depression [53, 58].

Perhaps the most appealing aspect of the risk-sensitive RL proposal is that it provides a simple and natural explanation for the central hallmark of depression, namely, the emotional disturbance, manifesting either as depressed mood or anhedonia [59]. As mentioned above, one important role of the RPE in temporal difference learning is to shape value estimates. While the basic mechanism for this is the same in standard temporal difference learning (see Eq. 3) and the risk-sensitive version (Eq. 4), in the latter case, when the learning rate for negative RPEs is larger than for positive ones ($\alpha^- > \alpha^+$), the learning process results, on average, in value estimates that are biased downward. If the environment is to any degree unpredictable, risk-sensitive RL will, by overweighting inferior outcomes, result in estimates of state value that are systematically pessimistic concerning the prospects for future reward. There is a good deal of intuitive appeal to the idea that depressed mood might constitute a direct expression of such pessimistic estimates of state value [60, 61]. Fortunately, there is also experimental evidence to support that intuition. Using experience sampling methods, Rutledge and colleagues [62, 63] have obtained empirical support for a model of mood dynamics that includes an expected value term, equivalent to the state value estimate in RL. High state values, within this model, contribute to an elevated mood, and low values to a depressed one. This proposal offers a direct link between risk-sensitive RL and the mood alteration in MDD: Asymmetric reinforcement learning systematically reduces state-value estimates, and the latter translate directly into depressed mood.

A further strength of the risk-sensitive RL proposal from Bennett and Niv [47] is that it explains anhedonia or depressed mood without assuming any disturbance in the hedonic impact of immediate reinforcers, a disturbance for which evidence has, perhaps surprisingly, been lacking in depression [64, 65, 66].

## Connecting the pieces: MDD and distributional coding

Risk-sensitive RL offers a promising account of the computational lesion, as it were, in depression. It does not, however, provide any insight into the disorder's pathophysiology at the biological level. The dropout theory, conversely, indicates a neurobiological problem without connecting this, in computational terms, to the clinical and behavioral manifestations of depression. This brings us to the third piece in our puzzle: recent work indicating that the VTA dopamine system employs a distributional code for value.

*Distributional reinforcement learning*, a paradigm widely explored in artificial intelligence [67], modifies temporal difference learning by replacing the scalar value typically used to represent the state value, instead using a vector that encodes a full probability distribution over state value. In effect, RPE and value signalling are both spread over a set of parallel channels, which vary along a spectrum from pessimism to optimism. Each channel's bias, on this spectrum, is determined by precisely the same learning-rate mechanism involved in risk-sensitive RL, specified above in Equation 4: Channels with a pessimistic bias arrive at this bias because they amplify negative over positive RPEs (i.e., $\alpha^- > \alpha^+$). Channels with an optimistic bias, conversely, do so because they amplify positive RPEs ($\alpha^+ > \alpha^-$). Distributional RL builds in a whole population of such channels, varying in their parameterization so as to span the full spectrum from pessimism to optimism. In statistical terms, each of these channels learns an estimate of one *expectile* of the value distribution (an analogue of the more familiar quantile; see [4, 68]). Pessimistic channels encode state values corresponding to lower expectiles of the distribution, while optimistic channels encode upper expectiles. Together, the entire collection of channels encodes the full probability distribution over future rewards.

In a recent experimental study, Dabney and colleagues [4] provided evidence that value representation in the brain may employ a distributional code very much akin to the one used in AI (see also [69]). As shown in Figure 3, dopamine neurons in VTA displayed strikingly different tuning curves in response to varying rewards, with some showing steeper slopes for positive RPEs, others steeper slopes for negative RPEs, with a smooth spectrum of asymmetries in between. Furthermore, detailed analysis of the same neuronsÕ reward responses indicated that different dopamine neurons were processing different estimates of state value, with neurons that amplified positive over negative RPEs receiving optimistic value estimates, and neurons amplifying negative RPEs more receiving pessimistic value estimates. This pattern makes sense computationally, if one notes that the parameters $\alpha^+$ and $\alpha^-$ in Equation 4 can be interpreted not only as learning rates, but also as multipliers of the RPE, in positive and negative domains respectively. The logic of risk-sensitive RL can then be seen to explain why dopamine neurons with asymmetric response functions, as in Figure 3, would drive learning of value estimates varying in their degree of optimism.

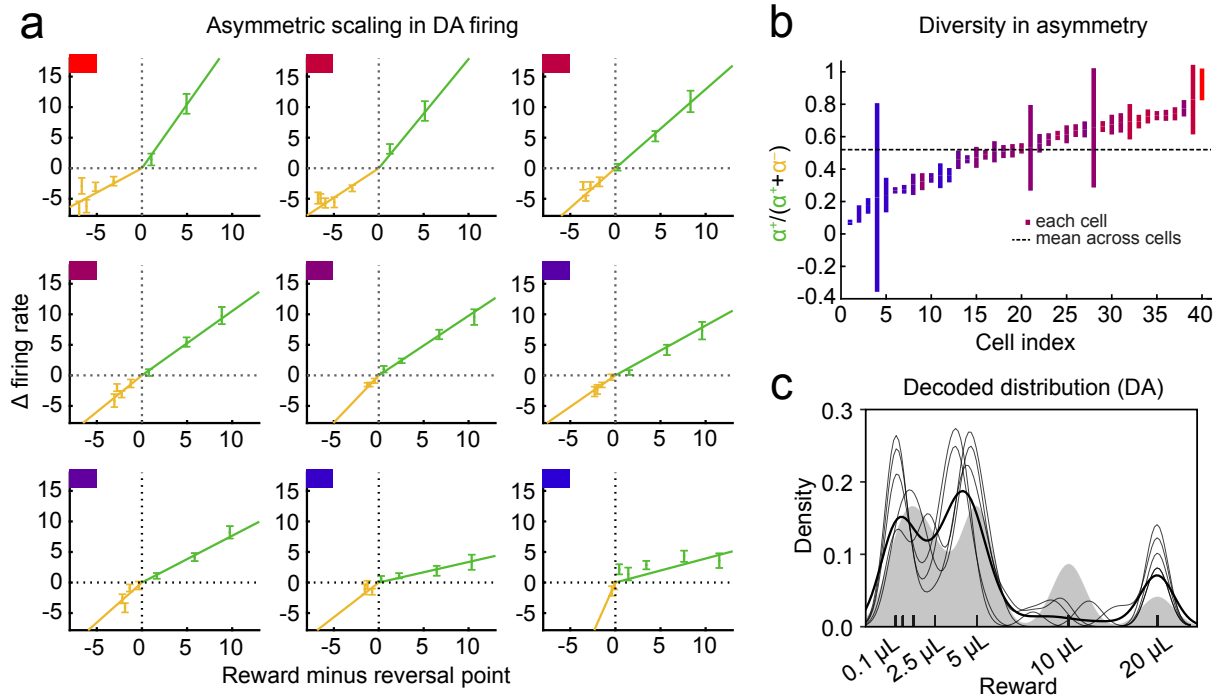

Figure 3: **Distributional RL.** Experimental results from Dabney et al. [4], consistent with distributional coding in the dopamine system. Behavioral task is the same as Figure 2a. **a.** Tuning curves for nine example VTA neurons, showing wide variation in the relative scaling of positive (green) and negative (yellow) RPE responses. 'Reversal point' refers to the reward magnitude (juice volume) that elicits neither an increase or decrease in firing rate, implying an RPE of magnitude zero. **b.** Summary of response asymmetries across a larger sample of recorded dopamine neurons. Quantities on the y axis indicate the direction and magnitude of the 'kink' in each neuron's response profile, with larger numbers indicating relatively steep slope in the positive RPE domain and smaller numbers a steeper slope in the negative domain. Color indicates the reversal point measured for that cell: red is optimistic and blue is pessimistic. **c.** Reconstruction of the ground-truth reward distribution from the dopaminergic population code. Grey indicates a smoothed version of the true distribution of rewards (juice volumes) in the experimental task. Thin lines show the inferred distribution, based on reward-triggered responses across a population of VTA neurons. (Estimates vary with initial random seed.) Thick line shows mean estimate across samples. All panels adapted from [4]
.

With these points in mind, we now return to the pathophysiology of depression. Recall that in the dropout theory, as proposed by Grace and colleagues [10], depression involves the inhibition of a subset of VTA dopamine neurons. Putting this together with the findings from Dabney and colleagues [4], we hypothesize that this inhibitory effect may predominantly affect *upper-expectile* neurons, that is, dopamine neurons with steeper response functions in the positive RPE domain than the negative (Figure 4). The consequence of this would be that the residual population of functionally active dopamine neurons would contain a preponderance of *lower-expectile* neurons, neurons with steeper response curves in the negative RPE domain. Computationally, the responses of these neurons are characterized by a parameterization in which $\alpha^-$ exceeds $\alpha^+$, driving the emergence of pessimistic state-value estimates. In this setting, the dropout described by Grace and colleagues [10] results in precisely the risk-sensitive RL scenario described by Bennett and Niv [47].

In Figure 4, we show with simulations that dropout of optimistic predictors in a distributional model directly explains several clinical and behavioral manifestations of MDD, including depressed mood or anhedonia, reduced RPE signalling in response to rewards, asymmetric learning effects, and risk aversion.

The most speculative component of this theory is the selectivity of the proposed dropout to optimistic VTA dopamine neurons. One possibility is optimistic and pessimistic neurons are differentially affected by pathophysiological processes. Recent evidence demonstrating optimistic and pessimistic neurons are biologically distinct suggests this is plausible. Striatal neurons preferentially expressing D1 receptors code for the optimistic side of the reward distribution, while D2-expressing cells code for the pessimistic side [70]. In a similar vein, it is worth noting the analogy to Parkinson's Disease, which of course differentially impacts specific sub-populations of midbrain dopaminergic neurons. The recent observation of anatomical gradients in distributional coding [71] also implies that selectivity in terms of optimism could be accounted for by selectivity in anatomical location. It is worth pointing out, for completeness, that distributional coding opens up other pathophysiological accounts that would have similar, if not identical consequences, such as for example an alteration in the tuning curves of individual dopaminergic neurons, without a wholesale suppression of their activity. However, given the compelling evidence provided by Grace and colleagues, a theory that anchors on their dropout theory seems particularly appealing.

## Steps toward completing the puzzle

One strong point of the theory we have proposed is that it translates into specific testable predictions. Our central conjecture is that depression is associated with a functional suppression of what we have referred to as upper-expectile neurons, that is, dopamine neurons with a steeper tuning curve for positive than negative RPEs. If this is accurate, then it should be manifest in the activity of dopamine neurons in depressed individuals. As discussed in the previous section and illustrated in Figure 3, Dabney and colleagues [4] established methods for measuring and comparing the tuning curves of dopamine neurons. If the technology existed to perform the same kinds of measurements in humans, the prediction to test would be clear: The distribution of slope asymmetries across dopamine neurons should be truncated, displaying an underrepresentation of upper-expectile neurons (see Figure 4). In the absence of techniques to perform such a test in humans, the same prediction can be pursued in the setting of an animal model of depression. In particular, the proposed experiment could be performed in rodents following CMS.

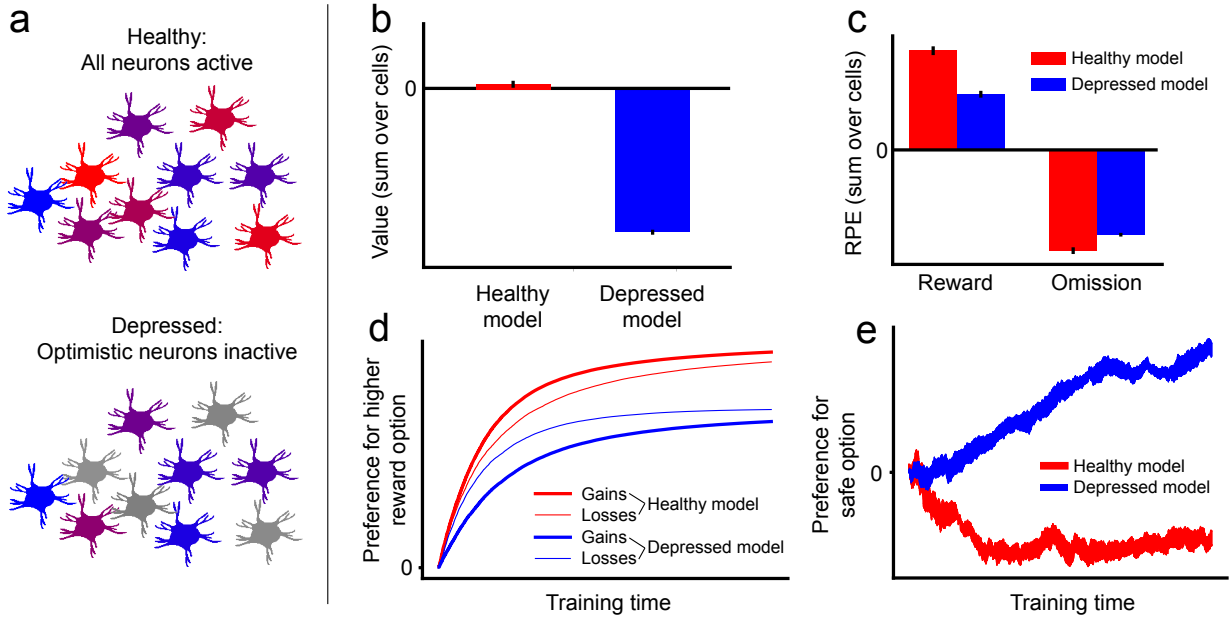

Figure 4: **Simulations of a distributional RL theory of depression. a.** Under the proposed theory, depression involves a selective silencing of upper-expectile (optimistic) dopamine neurons. Panels b-e show behavior of a standard distributional RL model (based on [4]) in simple, illustrative tasks. The 'healthy' model contains simulated dopamine neurons spanning a full range of expectiles. The 'depressed' model omits a set of units coding for upper expectiles. **b.** In a task yielding, on each trial, a gain or loss of $1 with even odds (a simplified version of a task format common in the literature, e.g., [37, 72]), the depressed model converges on lower estimates of state value, potentially accounting for depressed mood [62]. **c.** In the same task, the depressed model has smaller total RPE, especially when receiving better-than-expected outcomes. **d.** Depressed model learns faster from losses than from gains. There are two task conditions modeled here, each involving a two-armed-bandit problem (cf. tasks in [43, 73, 74]). In the 'Gains' condition, one action yields $1 and the other action $0; in the 'Losses' condition, one action wins $0 and the other action loses $1. Standard error is less than line thickness. **e.** Depressed model learns to be risk-averse. Here on each trial there is a choice between (1) a guaranteed win of 50¢ and (2) a gamble yielding, with even odds, either $1 or $0 (cf. tasks in [75, 76]).

If the foregoing central prediction can be confirmed, then a wide range of other questions would become relevant. To start, on the neurobiological level, if depression involves a selective suppression of upper-expectile dopamine neurons, what is the mechanism driving this selectivity? Some leverage might be gained on this question by further progress in mapping the diverse genetic and biochemical profiles of VTA neurons [77, 78, 79]. The differential D1/D2 receptor expression of optimistic versus pessimistic striatal neurons [70] suggests novel routes for testing our hypothesis. We predict that suppression of dopamine neurons induced by CMS [9, 10, 11] is specific to D1 circuits. Conversely, long-term inhibition of D1- but not D2-expressing striatal neurons, or the dopamine cells that project preferentially to them, should lead to depression-like symptoms. Of course, if insight can be gained into why upper-expectile neurons are specifically affected, that could conceivably translate into a better understanding of both the efficacy and limitations of current therapies, and perhaps to the development of strategic new therapeutic interventions, either through pharamacology [10, 15, 16], deep brain stimulation [20, 80], or other modalities. It is also of interest whether selective suppression arises from an evolutionarily adaptive mechanism heuristically tuning the organism's behavior to the statistics of its environment.

While we have emphasized the considerations that seem to recommend the proposed theory, it is of course also important to acknowledge a number of limitations and liabilities. First, while our theory goes beyond most current pathophysiological accounts of depression by tying neurobiological factors, through computation, to mood and behavior, it must be noted that existing experimental observations concerning decision-making behavior in depression are complicated, and effects such as risk aversion, impaired learning from positive outcomes, and accentuation of learning driven by aversive outcomes, while all frequently observed, have not been confirmed under all experimental conditions [32, 63, 81, 82]. Furthermore, there are some behavioral concomitants of depression that are not addressed by our theory, such as sleep disturbance [59] (although see [83]). At a finer grain, while our theory does provide an explanation for depressed mood, it does not directly explain why depressed individuals are prone to offer pessimistic predictions about specific future outcomes [60, 84, 85, 86, 87]. This finding may be reconciled with our theory by existing evidence for mood-congruence effects in cognition, whereby attention, memory and judgment are biased to align with preexisting emotional state [88, 89] (although see [90]), or, perhaps relatedly, by the observation that state values learned through incremental RL are integrated into prospective decisions [91, 92]. Another clinical aspect of depression that our theory does not directly address is disturbances in mental and physical effort, so-called motivational anhedonia [44]. However, studies on this aspect of MDD suggest that deficits in effort mobilization stem from a more primary disturbance in estimating anticipated payoffs of effort, thus affecting the cost-benefit analysis involved in effort-based decision making [44, 65, 93, 94], a possibility which dovetails well with the theory we have proposed. Further progress in understanding this aspect of depression might be gained by taking into account recent work seeking to understand the neural mechanisms underlying effort-based cost-benefit analyses in the normative setting [95].

## Conclusion

In this Perspective, we have assembled a set of observations from three different literatures, bridging between pathophysiology research, computational psychiatry, and basic neuroscience to construct a novel theory of depression. Like all current theories, the one we have proposed is incomplete, dependent on complex and in some cases contradictory data, and partially speculative. However, by highlighting some previously unexplored mechanistic possibilities, we are hopeful that the theory we have proposed may stimulate next steps in depression research that can yield a clearer picture of this devastating and common disorder. Among other things, the proposal we have offered provides an illustration of the opportunities that now exist to integrate across biological, computational and clinical perspectives, in the pursuit of a more comprehensive understanding of mental disorders.

**Code availability.** Simulation code will be open-sourced to align with publication.

**Data availability.** Not applicable.

**Acknowledgements.** We are grateful to Daniel Bennett and Nao Uchida for useful discussion.

**Author contributions.** All authors contributed equally.

**Competing Interests.** The authors declare no competing interests.